# Large Bandgap Opening Between Graphene Dirac Cones Induced by Na Adsorption onto an Ir Superlattice

Marco Papagno,[1] Stefano Rusponi,[2] Polina Makarovna Sheverdyaeva,[1] Sergio Vlaic,[2] Markus Etzkorn,[2] Daniela Pacilé,[1,3] Paolo Moras,[1] Carlo Carbone,[1] and Harald Brune[2]

[1] Istituto di Struttura della Materia, Consiglio Nazionale delle Ricerche, Trieste, Italy
[2] Institute of Condensed Matter Physics (ICMP), Station 3, Ecole Polytechnique Fédérale de Lausanne (EPFL), CH-1015 Lausanne, Switzerland
[3] Dipartimento Fisica, Università della Calabria, and INFN - Gruppo collegato di Cosenza, 87036 Dipartimento Fisica, Università della Calabria, and INFN (CS), Italy

**ABSTRACT** We investigate the effects of Na adsorption on the electronic structure of bare and Ir cluster superlattice covered epitaxial graphene on Ir(111) using angle-resolved photoemission spectroscopy and scanning tunneling microscopy. At Na saturation coverage a massive charge migration from sodium atoms to graphene raises the graphene Fermi level by about 1.4 eV relative to its neutrality point. We find that Na is adsorbed on top of the graphene layer and when coadsorbed onto an Ir cluster superlattice it results in the opening of a large bandgap of $\Delta_{Na/Ir/G}$ = 740 meV comparable to the one of Ge and with preserved high group velocity of the charge carriers.

Graphene (G) exhibits several exotic properties that can be ascribed to its two-dimensional character and band topology.[1-6] In native, freestanding graphene (Figure 1(a)), the π and π* states disperse linearly in the proximity of the Fermi level ($E_F$) and define the so-called Dirac cones, which are centered at the *K*-points of the reciprocal space and degenerate at the Dirac energy ($E_D$), where the cone vertices meet. This band structure, and the related high electron mobility, makes graphene appealing for nanoelectronics. Since graphene is a semiconductor with zero bandgap (Δ), the very first requisite for the exploitation of graphene for logic device applications consists in opening a bandgap between the π and π* states. Several methods exist to achieve this goal. Cutting graphene into stripes (graphene nano-ribbons) is an efficient way to break the degeneracy at $E_D$.[7-9] Recently,[10] the *bottom-up* approach has led to the production of nano-ribbons with uniform width and well-defined edges, however, the extension of this method to semiconductor substrates remains challenging. Alternatively, the tunable bandgap in gated graphene bi-layers ranging from zero to 250 meV may enable novel nanoelectronic and nanophotonic devices.[11] Angle-resolved photoemission spectroscopy (ARPES) experiments[12] show, however, that the band structure of these systems does not fulfill the requirement of high electron group velocity. Patterning,[13-15] adsorption[16-18] and intercalation of suitable elements,[19,20] strain[21,22] or substrate induced symmetry breaking[23-25] are other techniques inducing bandgaps in graphene layers ranging from a few tens to a few hundreds meV. In this context, metal substrates to which graphene couples weakly such as Pt(111)[26] and Ir(111),[27] offer interesting opportunities. Due to the large G/substrate spacing (about 3.7 Å) the band topology of the Dirac cones is weakly perturbed in these systems,[24,28] at variance with the case of strongly interacting metals (Ni(111),[29] Ru(0001)[30,31]), where the spacing is reduced to about 2.1 Å. Graphene is thus relatively well decoupled from the Ir(111) substrate and in fact the high group velocity of its charge carriers is preserved.[24] The lattice mismatch between graphene and Ir results in a long-period moiré superstructure, which weakly modulates the crystal potential and affects the electronic bands of graphene.[24,32,33] This results in the opening of a small bandgap of about 140 meV between the π and π* states at the *K*-point of the graphene Brillouin zone (see Figure 1(b)). These states can be further modified by ordered adsorption of atomic hydrogen[18] or Ir clusters.[24] H and Ir atoms selectively adsorb on the *hcp* regions of the moiré superstructure and promote a local re-hybridization from $sp^2$ to $sp^3$ bonding of the carbon orbitals.[34] The "floating" graphene layer becomes then a periodically bound layer and a bandgap ranging from 340 to 430 meV opens between π and π* states (Figure 1(c)).[24]

In this work we show a new method to further increase the bandgap between the Dirac cones in adsorbed graphene. We demonstrate that Na adsorption on bare (Na/G) and on a well-ordered superlattice of Ir clusters (Na/Ir/G) on epitaxial graphene on Ir(111) produces very large bandgaps while preserving the shape of the Dirac cone and the high electron group velocity. For Na/G (Figure 1(d)) we observe a bandgap of $\Delta_{Na/G}$ = 320 meV, whereas in the case of Na/Ir/G it reaches $\Delta_{Na/Ir/G}$ = 740 meV (see Figure 1(e)), which is very close to the bandgap of conventional semiconductors. Direct access to the $\pi$ - $\pi^*$ bandgap for photoemission is made possible by the strong charge migration from Na toward graphene, which partly fills the $\pi^*$ state. Scanning tunneling microscopy (STM) measurements show that Na does not intercalate at room temperature (RT), it forms an adsorbed layer commensurate to the graphene moiré structure. The effect of the Na layer is attributed to a strengthening of the graphene sublattice asymmetry induced by the moiré superstructure. The coadsorption with Ir clusters further enhances the difference between the two carbon sublattices explaining the large bandgap.

**RESULTS AND DISCUSSION**

Figures 2(a) and 2(b) show ARPES maps of the electronic band dispersion of Na/G and Na/Ir/G along the direction perpendicular to ΓK, that we label pΓK (see the red line in the inset of Figure 2(a)). Due to strong electron doping the $\pi^*$ state is shifted below $E_F$ as compared with bare graphene on Ir(111). The charge transfer from sodium atoms to graphene estimated from the relative size of graphene Fermi surface with respect to the surface Brillouin zone of graphite is 0.032 (0.028) electrons per C atom, equivalent to 0.26 (0.22) electrons per Na atom for Na/G (Na/Ir/G). From the constant $dE/dk$ gradient of the band dispersion of the $\pi^*$ state we evaluate the electron group velocities to $v_{Na/G}$ ≈ 5.5 eV Å and $v_{Na/Ir/G}$ ≈ 5.2 eV Å, respectively. These values are very close to the ones obtained by third nearest neighbor tight-binding calculations and in magneto-transport measurements of freestanding graphene.[35] For both systems, the $\pi^*$ ($\pi$) state disperses linearly downward (upward) and reaches a minimum (maximum) at the graphene $K$-point. There, $\pi$ and $\pi^*$ states open up a bandgap of $\Delta_{Na/G}$ = 320 meV for Na/G and of $\Delta_{Na/Ir/G}$ = 740 meV for the Na/Ir/G surface. The energy distribution curves (EDCs) along the ΓK direction are shown in Figures 2(c) and 2(d) and reveal broad line-shapes, which cause an overlap of the intensity tails from the top of the valence band and the bottom of the conduction band. Nevertheless, the dispersion of the bands and the electronic gap between the $\pi$ and $\pi^*$ states are well resolved.

The graphene $\pi^*$ state exhibits a kink at about 180 meV below $E_F$ highlighted by arrows in Figures 2(a) and 2(b). Similarly to K-doped graphene/Ir,[36] this feature can be interpreted as a renormalization of the Dirac cone band introduced by electron-phonon interactions, however, the overlap between Ir 5$d$ and G $\pi$ states near $E_F$[33] does not allow a unequivocal assignment.

Extended EDCs integrated over 6° around the *K*-point in Figure 3(a) allow quantifying the shift of graphene bands. In particular, from the displacement of the σ band we measure a downward shift of 1.5 eV for Na/G (blue full curve) and 1.3 eV for Na/Ir/G (blue dashed curve) as compared to the undoped surfaces displayed in red. These values are in good agreement with experimental observations on K-doped graphene on Ir(111),[36] and KC$_8$,[37] and with the theoretical prediction for Li-doped graphene.[38]

Figures 4(a) and 4(b) show STM images of the bare and partly Na covered graphene, respectively demonstrating that Na does not intercalate at RT but rather decorates the step edges and forms single layer islands with lateral dimensions that vary from 10 nm to 100 nm for a coverage of θ ~ 0.5 ML.[39] Note that the adsorption properties of Na on graphite significantly differ from the other alkali metals, Na does not form intercalation compounds while the other alkalis do.[40] This singular behavior seems to be present also on metal supported graphene. In contrast to Na, K has been reported to penetrate the graphene layer on Ni(111) already at RT and to act as spacer.[41,42] In these systems the increased distance reduces the hybridization between Ni 3*d* and C 2$p_z$ states and forces the dispersion of the π band to turn from parabolic to linear.

For our system we observe that the Na island size increases with exposure time until the whole graphene surface is covered for θ = 1 ML. Alkali metals on graphite possess a "correlated liquid" phase in the low-coverage regime. This phase originates from the charge transfer from the alkalis to the graphite leading to vertical dipoles which repel each other preventing the formation of islands and maximizing the adatom-adatom distance.[43,44] As the alkali coverage reaches a critical density, a nucleation to a more closely packed configuration is observed, which results in the formation of alkali islands almost decoupled from the substrate. In our case, we observe island formation from lowest coverage on, reminiscent of an attractive interaction between Na adatoms. In analogy with K[36,45] and Li[38,46] on graphene, Na is expected to form a (2×2) or (√3×√3)R30° phase. We were unable to reveal the Na superstructure, presumably due to the weakness of the Na-C bond with respect to interaction with the STM tip often resulting in tip-induced displacements of Na atoms. However, for particular tunneling conditions, the Na layer appears transparent and the moiré pattern of the underlying graphene with attenuated corrugation appears in the STM images revealing that Na adsorption does not change the graphene moiré structure. We further find that the edges of the Na islands are always aligned to those of the underlying graphene moiré superstructure, thus suggesting a Na superstructure commensurate with that of the graphene moiré. From the fact that the moiré unit cell contains close to (10x10) graphene unit cells above a (9x9) Ir(111) unit cell,[27] we conclude a (2x2)-Na superstructure.

Further support for the (2×2) Na/G phase is provided by core level measurements. In Figure 3(b) we report the C 1*s* peak for bare graphene (red full curve) centered at -284.15 eV below $E_F$, in agreement with earlier results.[47,48] Following Na adsorption (full blue curve) we observe a shift of the main peak by 1.05 eV toward higher binding energy and a shoulder at ~ -286 eV. The same splitting has been observed for (2×2) K/graphite and assigned to energy losses associated with the metallic overlayer.[49] We therefore suggest that this shoulder is due to energy losses associated with (2x2) Na.

The presence of Ir clusters on graphene induces small changes to the lineshape of the C 1*s* resonance (dashed red curve in Figure 3(b)). Na deposition on the cluster superlattice (dashed blue curve) also produces a downward shift of the C 1*s* peak, which is however reduced by ~200 meV with respect to the shift observed in Na/G. We attribute this effect to the smaller saturation coverage of sodium at the Ir/G surface caused by site blocking. Similar lineshapes of the photoemission spectra for the Na covered surface, with and without the cluster superlattice, suggest that, for the sodium saturation coverage, the presence of Ir clusters only weakly affects the adsorption properties of Na.

In graphene supported on Ni(111) and in graphite, alkali metal intercalation is supposed to happen at surface defects such as step edges or walls between different rotational domains.[41, 50] In our case, the high quality of the graphene layer with and without the Ir clusters may prevent Na intercalation. Figures 5(a) and 5(b) show constant-energy ARPES maps for G and Ir/G, respectively. The G/Ir(111) moiré superpotential[32] induces six replicas of the π state, labeled "1-6", surrounding the main Dirac cone, "0", centered at the *K*-point. The Ir cluster superlattice introduces anisotropies in the spectral-weight distribution by suppressing the photoemission intensity of three of the six replicas, but it does not affect the quality of the graphene layer.[24] The absence of additional Dirac cones rotated away from the *K*-points also rules out rotational domains[33] in both surfaces.

We propose two possible origins of the observed bandgaps. First, Na adatoms modify the crystal potential, thus $\Delta_{Na/G}$ is expected to be different from that of undoped graphene. Previous photoemission experiments of Na on highly oriented pyrolytic graphite revealed the presence of an energy gap in the surface electronic structure.[51] This bandgap has been attributed to an inhomogeneous charge transfer through the stack of the graphene layers, which perturbs the symmetry of the graphite lattice. Similarly, we may argue in the present case that, upon deposition of Na, the electronic charge is differently distributed between the graphene overlayer and the metallic substrate modifying the symmetry of the system.

The second possibility is related to the commensurate phase of Na adatoms with respect to the graphene moiré. It has been found theoretically[38] that a (2×2) superstructure of any adsorbate on an unperturbed graphene layer should not open a bandgap, as the (2×2) phase does not break the degeneracy of the two carbon sublattices, and in fact no gap has been experimentally reported on any (2×2) phase on graphene on SiC.[6,45,52] However, the graphene moiré on Ir(111) breaks this degeneracy for 2/3 of the graphene layer where either one of the two C atoms is adsorbed on top an Ir atom whereas the other is located on a three-fold hollow site.[27] Accordingly, there is a small electronic gap at the $K$-point.[24,32,33] The as grown (2×2)-Na superstructure can enhance this difference between the C sublattices and thus increase the bandgap. The C sublattice degeneracy can also be lifted through Ir clusters due to the peculiar properties of the Ir adatoms to nucleate only on the *hcp* sites of the moiré and due to the local re-hybridization of the C atoms.[27,34] The combination of Ir clusters and Na atoms induces the strongest breaking of the C sublattice degeneracy and leads to the large bandgap of $\Delta_{Na/Ir/G}$ =740 meV.

**CONCLUSION**

We have demonstrated that Na adsorption onto bare and Ir-cluster superlattice pre-covered epitaxial graphene on Ir(111) produces very large bandgaps between the $\pi$ and $\pi^*$ states amounting to $\Delta_{Na/G}$ =320 meV and $\Delta_{Na/Ir/G}$ = 740 meV, respectively. The latter is comparable with that of Ge. In addition, we find that the linear band dispersion is only marginally perturbed giving rise to electron group velocities close to the ones of freestanding graphene. Photoemission experiments show that there is a substantial charge transfer from Na to graphene which promotes a downward shift of electronic states by more than 1 eV. STM data show that Na is adsorbed on top of graphene, most likely in a (2×2) phase, and does not intercalate at RT. Though a (2×2) reconstruction is not expected to remove the degeneracy between graphene states at the $K$-point, a bandgap may be induced by a hybridization between the electron wave functions with wave vectors $K$ and $K$' promoted by the translation symmetry breaking of the combined Na-graphene-Ir(111) lattice. We attribute the wide $\Delta_{Na/Ir/G}$ bandgap to a mutual phenomenon of Na and Ir which breaks the degeneracy of the two sublattices of graphene. This combined effect may apply to systems other than graphene on Ir. The prerequisites for our method to work are a graphene moiré pattern acting as template for the growth of nanostructure superlattices and an ordered array of dopant that does not intercalate. Therefore a transfer to SiC is in principle feasible, since also this system exhibits a moiré pattern and alkalis do not intercalate at RT.[6,45,46] Our method has the advantage to open large bandgaps in the electronic states while preserving the appealing transport properties of graphene.

**METHODS**

The ARPES experiments were performed at the VUV-Photoemission beamline at Elettra in Trieste and the STM experiments at EPFL under ultrahigh vacuum (UHV) condition with a base pressure better than $5\times10^{-11}$ mbar. The Ir(111) crystal was cleaned by repeated cycles of Ar$^+$ sputtering ($E = 1.2$ keV) and annealing at $T = 1500$ K. Order and cleanness of the sample were monitored by low-energy electron diffraction and photoemission spectroscopy in Trieste and by STM in Lausanne. Graphene was grown by ethylene ($C_2H_4$) exposure at a pressure of $2.5\times10^{-6}$ mbar for 53 sec. (corresponding to 100 L) with the sample temperature held at 1300 K.[24,53] Ir was evaporated from a current heated thin film plate (0.5 mm width, 0.1 mm thickness). Evaporation rate (about $2.0\times10^{-4}$ ML/s) and coverage were determined by core level measurements. The calibration was crosschecked with STM. Iridium clusters were grown by evaporating 0.15 ML Ir [for Ir, one monolayer, ML, is defined as one atom per Ir substrate atom] on the graphene surface held at a temperature of 350 K. Under these experimental conditions all the *hcp* regions of the graphene moiré are covered with Ir clusters.[27] Na was evaporated from a commercial getter source (SAES) at room temperature. Angle-resolved photoemission measurements were carried out for the alkali-metal adsorbed systems at saturation coverage and immediately after deposition. Photoemission experiments were carried out at 120 eV photon energy with an energy resolution of 30 meV and at 120 K. This photon energy corresponds to the "Cooper minimum" in the photoemission cross-section of the Ir 5*d* states, and allows to minimize the number of electrons photoemitted from the *d* states of the metal relative to those from the π states of graphene. ARPES data were collected using a Scienta R4000 electron energy analyzer, which allows spectra to be recorded simultaneously within an angular aperture of 30º. ARPES maps were acquired by spanning the azimuthal angle over an angular range of more than 70º by step of 0.5º. STM measurements were carried out with a homebuilt variable temperature scanning tunneling microscope at 120 K.[54]

*Acknowledgments.* The authors thank M. Polini and L. Moreschini for helpful discussions. This work has been supported by the European Science Foundation (ESF) under the EUROCORES

**REFERENCES AND NOTES**

1. Berger, C.; Song, Z.; Li, X.; Wu, X.; Brown, N.; Naud, C.; Mayou, D.; Li, T.; Hass, J.; Marchenkov, A. N.; *et al.* Electron Confinement and Coherence in Patterned Epitaxial Graphene. *Science* **2006**, 312, 1191-1196.

2. Novoselov, K. S.; Geim, A. K.; Morosov, S. V.; Jiang, D.; Zhang, Y.; Dubonos, S. V.; Grigorieva, I. V.; Firsov, A. A. Electric Field Effect in Atomically Thin Carbon Films. *Science* **2004**, 306, 666-669.

3. Novoselov, K. S.; Geim, A. K.; Morozov, S. V.; Jiang, D.; Katsnelson, M. I.; Grigorieva, I. V.; Dubonos, S. V.; Firsov, A. A. Two-dimensional Gas of Massless Dirac Fermions in Graphene. *Nature* **2005**, 438, 197-200.

4. Zhang, Y.; Tan, Y. W.; Stormer, H. L.; Kim, P. Experimental Observation of the Quantum Hall Effect and Berry's Phase in Graphene. *Nature* **2005**, 438, 201-204.

5. Novoselov, K. S.; Jiang, Z.; Zhang, Y.; Morozov, S. V.; Stormer, H. L.; Zeitler, U.; Maan, J. C.; Boebinger, G. S.; Kim, P.; Geim, A. K. Room-Temperature Quantum Hall Effect in Graphene. *Science* **2007**, 315, 1379.

6. Bostwick, A.; Ohta, T.; Seyller, T.; Horn, K.; Rotenberg, E. Quasiparticle Dynamics in Graphene. *Nat. Phys.* **2007**, 3, 36-40.

7. Yang, L.; Park, C.-H.; Son, Y.-W.; Cohen, M. L.; Louie, S. G. Quasiparticle Energies and Band Gaps in Graphene Nanoribbons. *Phys. Rev. Lett.* **2007**, 99, 186801-186804.

8. Han, M. Y.; Özyilmaz, B.; Zhang, Y.; Kim, P. Energy Band-Gap Engineering of Graphene Nanoribbons. *Phys. Rev. Lett.* **2007**, 98, 206805-206809.

9. Kim, P. in *Tech. Dig. IEDM* **2009**, 241-244 (IEEE).

10. Cai, J.; Ruffieux, P.; Jaafar, R.; Bieri, M.; Braun, T.; Blankenburg, S.; Muoth, M.; Seitsonen, A. P.; Saleh, M.; Feng, X.; *et al.* Atomically Precise Bottom-Up Fabrication of Graphene Nanoribbons. *Nature* **2010**, 466, 470-473.

11. Zhang, Y.; Tang, T.-T.; Girit, C.; Hao, Z.; Martin, M. C.; Zettl, A.; Crommie, M.; Shen, Y. R.; Wang, F. Direct Observation of a Widely Tunable Bandgap in Bilayer Graphene. *Nature* **2009**, 459, 820-823.

12. Ohta, T.; Bostwick, A.; Seyller, Th.; Horn, K.; Rotenberg, E. Controlling the Electronic Structure of Bilayer Graphene. *Science* **2006**, 313, 951-954.

13. Liang, X.; Jung, Y.-S.; Wu, S.; Ismach, A.; Olynick, D. L.; Cabrini, S.; Borok, J. Formation of Bandgap and Subbands in Graphene Nanomeshes with Sub-10 nm Ribbon Width Fabricated via Nanoimprint Lithography. *Nano Lett.* **2010**, 10, 2454-2460.

14. Bai, J.; Zhong, X.; Jiang, S.; Huang, Y.; Duan, X. Graphene Nanomesh. *Nat. Nanotechnol.* **2010**, 5, 190-194.

15. Pedersen, T. G.; Flindt, C.; Pedersen, J.; Mortensen, N. A.; Jauho, A.-P.; Pedersen, K. Graphene Antidot Lattices: Designed Defects and Spin Qubits. *Phys. Rev. Lett.* **2008**, 100, 136804-136808.

16. Yavari, F.; Kritzinger, C.; Gaire, C.; Song, L.; Gulapalli, H.; Borca-Tasciuc, T.; Ajayan, P. M.; Koratkar, N. Tunable Bandgap in Graphene by the Controlled Adsorption of Water Molecules. *Small* **2010**, 6, 2535-2538.

17. Park, C. –H.; Louie, S. G. Energy Gaps and Stark Effect in Boron Nitride Nanoribbons *Nano Lett.* **2008**, 8, 2200-2203.

18. Balog, R.; Jørgense, B.; Nilsson, L.; Andersen, M.; Rienks, E.; Bianchi, M.; Fanetti, M.; Lægsgaard, E.; Baraldi, A.; Lizzit, S. *et al.* Bandgap Opening in Graphene Induced by Patterned Hydrogen Adsorption. *Nat. Mater.* **2010**, 9, 315-319.

19. Haberer, D.; Vyalikh, D. V.; Taioli, S.; Dora, B.; Farjam, M.; Fink, J.; Marchenko, D.; Pichler, T.; Ziegler, K.; Simonicci, S.; *et al.* Tunable Band Gap in Hydrogenated Quasi-Free-Standing Graphene. *Nano Lett.* **2010**, 10, 3360-3366.

20. Varykhalov, A.; Scholz, M. R.; Kim, T. K.; Rader, O. Effect of Noble-Metal Contacts on Doping and Band Gap of Graphene. *Phys. Rev. B* **2010**, 82, 121101-121104.

21. Ni, Z. H.; Yu, T.; Lu, Y. H.; Wang, Y. Y.; Feng, Y. P.; Shen, Z. X. Uniaxial Strain on Graphene: Raman Spectroscopy Study and Band-Gap Opening. *ACS Nano* **2008**, 11, 2301-2305.

22. Guinea, F.; Katsnelson, M. I.; Geim, A. K. Energy Gaps and a Zero-Field Quantum Hall Effect in Graphene by Strain Engineering. *Nat. Phys.* **2010**, 6, 30-33.

23. Zhou, S. Y.; Gweon, G.-H.; Federov, A. V.; First, P. N.; De Heer, W. A.; Lee, D.-H.; Guinea, F.; Catro Neto, A. H.; Lanzara, A. Substrate-Induced Bandgap Opening in Epitaxial Graphene. *Nat. Mater.* **2007,** 6, 770-775.

24. Rusponi, S.; Papagno, M.; Moras, P.; Vlaic, S.; Etzkorn, M.; Shverdyaeva, P. M.; Pacilè, D.; Brune, H.; Carbone, C. Highly Anisotropic Dirac Cones in Epitaxial Graphene Modulated by an Island Superlattice. *Phys. Rev. Lett.* **2010**, 105, 246803-246806.

25. Oshima, C.; Nagashima, A. Ultra-Thin Epitaxial Films of Graphite and Hexagonal Boron Nitride on Solid Surfaces. *J. Phys.:Condens. Matter* **1997**, 9, 1-20.

26. Zi-pu. H.; Ogletree, D. F.; Van Hove, M. A.; Somorjai, G. A. Leed Theory for Incommensurate Overlayers: Application to Graphite on Pt(111). *Surf. Sci* **1987**, 180, 433-459.

27. N'Diaye, A. T.: Bleikamp, S.; Feibelman, P. J.; Michely, T. Two-Dimensional Ir cluster Lattice on a Graphene Moiré on Ir(111). *Phys. Rev. Lett.* **2006**, 97, 215501-215503.

28. Sutter, P.; Sadowski, J. T.; Sutter, E. Graphene on Pt(111): Growth and Substrate Interaction. *Phys. Rev. B* **2009**, 80, 245411-245420.

29. Gamo, Y.; Nagashima, A.; Wakabayashi, M.; Terai, M.; Oshima, C. Atomic Structure of Monolayer Graphite Formed on Ni(111). *Surf. Sci.* **1997**, 374, 61-64.

30. Moritz, W.; Wang, B.; Bocquet, M.-L.; Brugger, T.; Greber, T.; Wintterlin, J.; Günther, S. Structure Determination of the Coincidence Phase of Graphene on Ru(0001). *Phys. Rev. Lett.* **2010**, 104, 136102-136105.

31. Stradi, D.; Barja, S.; Díaz, C.; Garnica, M.; Borca, B.; Hinarejos, J. J.; Sánchez-Portal, D.; Alcamí, M.; Arnau, A.; Vázquez de Parga, A. L.; Miranda, R. *et al.* Role of Dispersion Forces in the Structure of Graphene Monolayers on Ru Surfaces. *Phys. Rev. Lett.* **2011**, 106, 186102-186105.

32. Pletikosić, I.; Kralj, M.; Pervan, P.; Brako, R.; Coraux, J.; N'Diaye, A. T.; Busse, C.; Michely, T. Dirac Cones and Minigaps for Graphene on Ir(111). *Phys. Rev. Lett.* **2009**, 102, 056808-056811.

33. Starodub, E.; Bostwick, A.; Moreschini, L.; Nie, S.; Gabaly, F.; McCarty, K. F.; Rotenberg, E. In-plane Orientation Effects on the Electronic Structure, Stability, and Raman Scattering of Monolayer Graphene on Ir(111). *Phys. Rev. B* **2011**, 83, 125428-125436.

34. Feibelman, P. J. Pinning of Graphene to Ir(111) by Flat Ir dots. *Phys. Rev. B* **2008**, 77, 165419-165425.

35. Reich, S.; Maultzsch, J.; Thomsen, C.; Ordejón, P. Tight-Binding Description of Graphene. *Phys. Rev. B* **2002**, 66, 035412-035416.

36. Bianchi, M.; Rienks, E. D.; Lizzit, S.; Baraldi, A.; Balog, R.; Hornekaer, L.; Hofmann, Ph. Electron-Phonon Coupling in Potassium-Doped Graphene: Angle-resolved Photoemission Spectroscopy. *Phys. Rev. B* **2010**, 81, 041403-041406 (R).

37. Grüneis, A.; Attaccalite, A.; Rubio, A.; Vyalikh, D.V.; Molodtsov, S. L.; Fink, J.; Follath, R.; Eberhrdt, W.; Pichler, T. Electronic Structure and Electron-Phonon Coupling of Doped Graphene layers in KC8. *Phys. Rev. B* **2009**, 79, 205106-205114.

38. Farjam, M.; Rafii-Tabar, H. Energy Gap Opening in Submonolayer Lithium on Graphene: Local Density Functional and Tight-Binding Calculations. *Phys. Rev. B* **2009**, 79, 045417-045423.

39. The Na coverage is defined with respect to a saturated ML.

40. Johnson, M. T.; Starnberg, H. I.; Hughes, H. P. Electronic Structure of Alkali Metal Overlayers on Graphite. *Surf. Sci.* **1986**, 178, 290-299.

41. Nagashima, A.; Tejima, N.; Oshima, C. Electronic States of the Pristine and Alkali-Metal-Intercalated Monolayer Graphite/Ni(111) Systems. *Phys. Rev. B* **1994**, 50, 17487-17495.

42. Grüneis, A.; Vyalikh, D.V. Tunable Hybridization Between Electronic States of Graphene and a Metal Surface. *Phys. Rev. B* **2008**, *77*, 193401-193404.

43. Hunt, M. R. C.; Palmer, R. E. The Development of Metallic Behaviour in Clusters on Surfaces. *Philos. Trans. R. Soc. London A* **1998**, 356, 231-247.

44. Yin, F.; Akola, J.; Koskinen, P.; Mannine, M.; Palmer, R. E. Bright Beaches of Nanoscale Potassium Islands on Graphite in STM Imaging. *Phys. Rev. Lett.* **2009**, 102, 106102-106105.

45. McChesney, J. L.; Bostwick, A.; Otha, T.; Seyller, T.; Horn, K.; González, J.; Rotenberg, E. Extended van Hove Singularity and Superconducting Instability in Doped Graphene. *Phys. Rev. Lett.* **2010**, 104**,** 1368039-136807.

46. Virojanadara, C.; Watcharinyanon, S.; Zakharov, A. A.; Johansson, L. I. Epitaxial Graphene on 6H-SiC and Li Intercalation. *Phys. Rev. B* **2010,** 82, 205402-205407.

47. Lizzit, S.; Zampieri, G.; Petaccia, L.; Larciprete, R.; Lacovig, P.; Rienks, E. D. L.; Bihlmayer, G.; Baraldi, A.; Hofmann, P. Band Dispersion in the Deep 1s Core Level of Graphene. *Nat. Phys.* **2010**, 6, 345-349.

48. Preobrajenski, A. B.; Ng, M. L.; Vinogradov, A. S.; Mårtensson, N. Controlling Graphene Corrugation on Lattice-Mismatched Substrates. *Phys. Rev. B* **2008**, 78, 073401-073404.

49. Bennich, P.; Puglia, C.; Brühwiler, P. A.; Nilsson, A.; Maxwell, A. J.; Sandell, A.; Mårtensson, N.; Rudolf, P. Photoemission Study of K on Graphite. *Phys. Rev. B* **1999**, 59, 8292-8304.

50. Wu, N. J.; Ignatiev, A. Potassium Absorption Into the Graphite(0001) Surface: Intercalation. *Phys. Rev. B* **1983**, 28, 7288–7293.

51. Pivetta, M.; Patthey, F.; Barke, I.; Hövel, H.; Delley, B.; Schneider, W. D. Gap Opening in the Surface Electronic Structure of Graphite Induced by Adsorption of Alkali Atoms: Photoemission Experiments and Density Functional Calculations. *Phys. Rev. B* **2005**, 71, 165430-165433.

52. Hwang, C. G.; Shin, S. Y.; Choi, S.-M.; Kim, N. D.; Uhm, S. H.; Kim, H. S.; Hwang, C. C.; Noh, D. Y.; Jhi, S.-H.; Chung, J. W. Stability of Graphene Band Structures Against an External Periodic Perturbation: Na on Graphene. *Phys. Rev. B* **2009**, 79, 115439-115443.

53. N'Diaye, A. T.; Coraux, J.; Plasa, T.; Busse, C.; Michely, T. Structure of epitaxial graphene on Ir(111). *New J. Phys.* **2008**, 10, 043033-043047.

54. Lehnert, A.; Buluschek, P.; Weiss, N.; Giesecke, J.; Treier, M.; Rusponi, S.; Brune H. High Resolution In Situ Magneto-Pptic Kerr Effect and Scanning Tunneling Microscopy Setup with all Optical Components in UHV. *Rev. Sci. Instrum.* **2009**, 80, 023902-023908.

**Figure captions**

Figure 1 Side view sketch of the atomic structure and of the π and π* bands dispersion close to the *K*-point of graphene Brillouin zone and to the Fermi level $E_F$ for (a) freestanding graphene (b) G (c) Ir/G (d) Na/G and (e) Na/Ir/G.

Figure 2(a) ARPES map along the pΓK direction for Na/G and (b) Na/Ir/G. The inset shows the 2D Brillouin zone of graphene. The red line indicates the pΓK direction. Black crosses overlaid to the ARPES maps mark the maximum of the photoemission spectral-weight intensity whereas arrows highlight kinks in the electronic band dispersion. (c) EDCs along the ΓK direction from $k = 1.4$ Å$^{-1}$ (bottom) to $k = 2.2$ Å$^{-1}$ (top) for Na/G and (d) Na/Ir/Na. Red curves are measured at the *K*-point.

Figure 3(a) Extended EDCs integrated over 6º around the *K*-point for G (solid red curve), Na/G (solid blue curve), Ir/G (dashed red curve), and Na/Ir/G (dashed blue curve). Na adsorption shifts the valence band by 1.5 eV (1.3 eV) in Na/G (Na/Ir/G) in respect to G (Ir/G) as estimated by the shift of the σ state. (b) C1*s* photoelectron spectra measured at normal emission with photon energy of 450 eV for G (solid red curve), Ir/G (dashed red curve), Na/G (solid blue curve), and Na/Ir/G (dashed blue curve).

Figure 4(a) Constant current STM image showing the moiré pattern of clean graphene on Ir(111) (tunnel voltage applied to sample $V_t = 0.05$ V; tunnel current $I_t = 10$ nA; size 72 nm *x* 72 nm). (b) Graphene covered by Na patches for a Na coverage Θ ≈ 0.5 ML ($V_t = -2.5$ V; $I_t = 10$ nA; size 56 nm *x* 56 nm).

Figure 5 Constant-energy ARPES maps at (a) $E-E_F = 400$ meV for G and (b) $E-E_F = 530$ meV Ir/G. Dashed line marks the first graphene Brillouin zone. Main Dirac cone is labeled "0", whereas "1-6" are replicas. In (b) only three out of six replicas are observed.

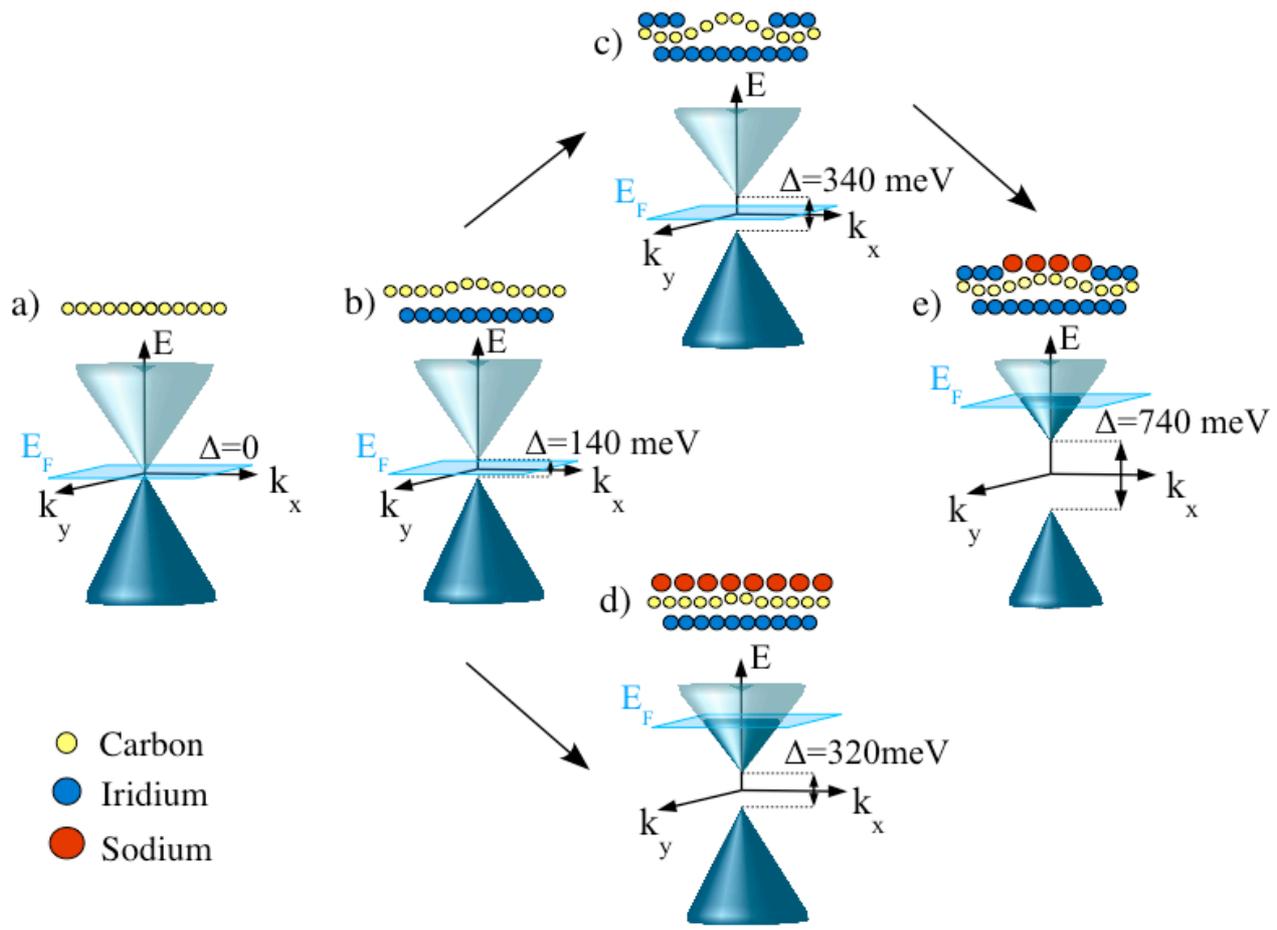

Figure 1

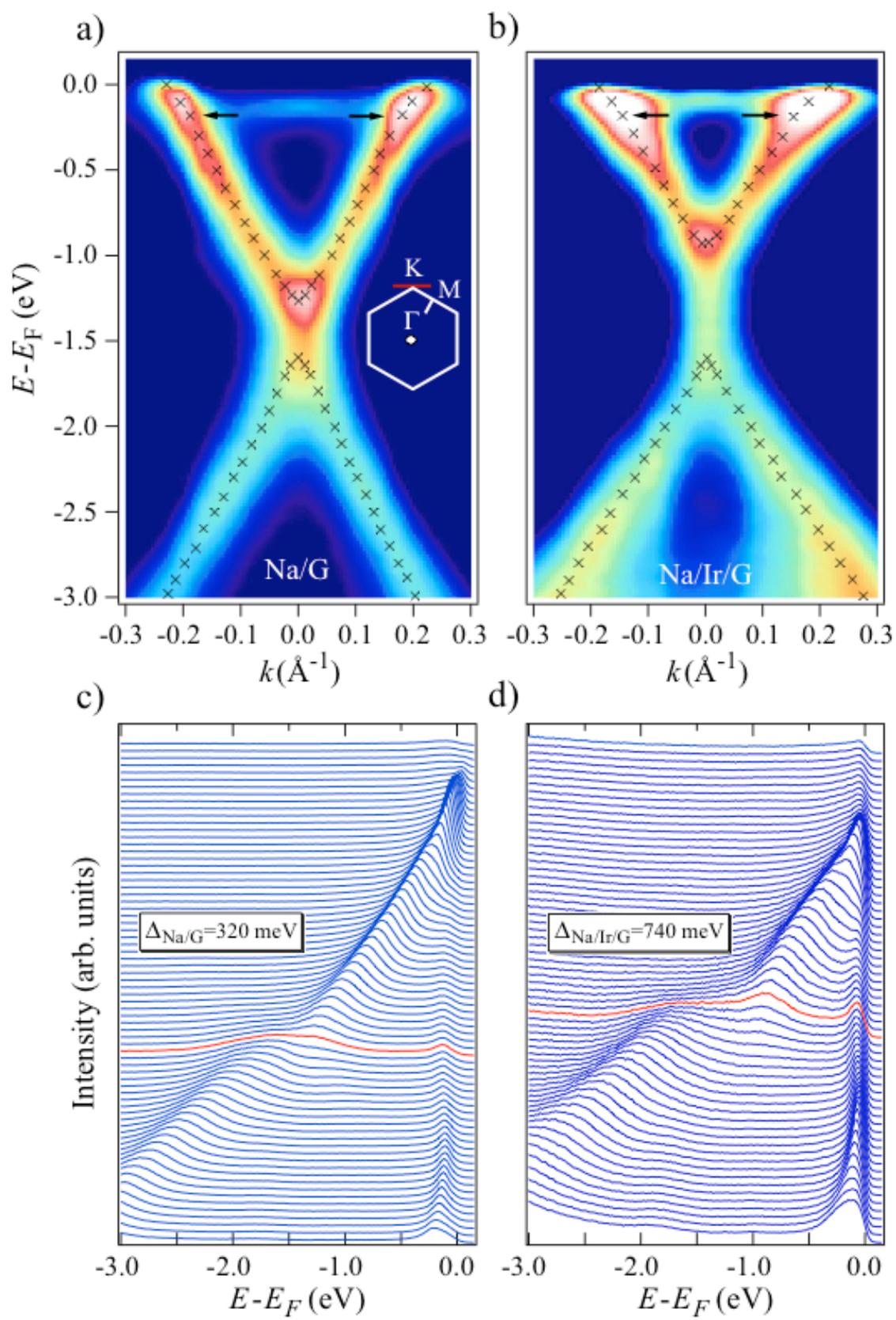

Figure 2

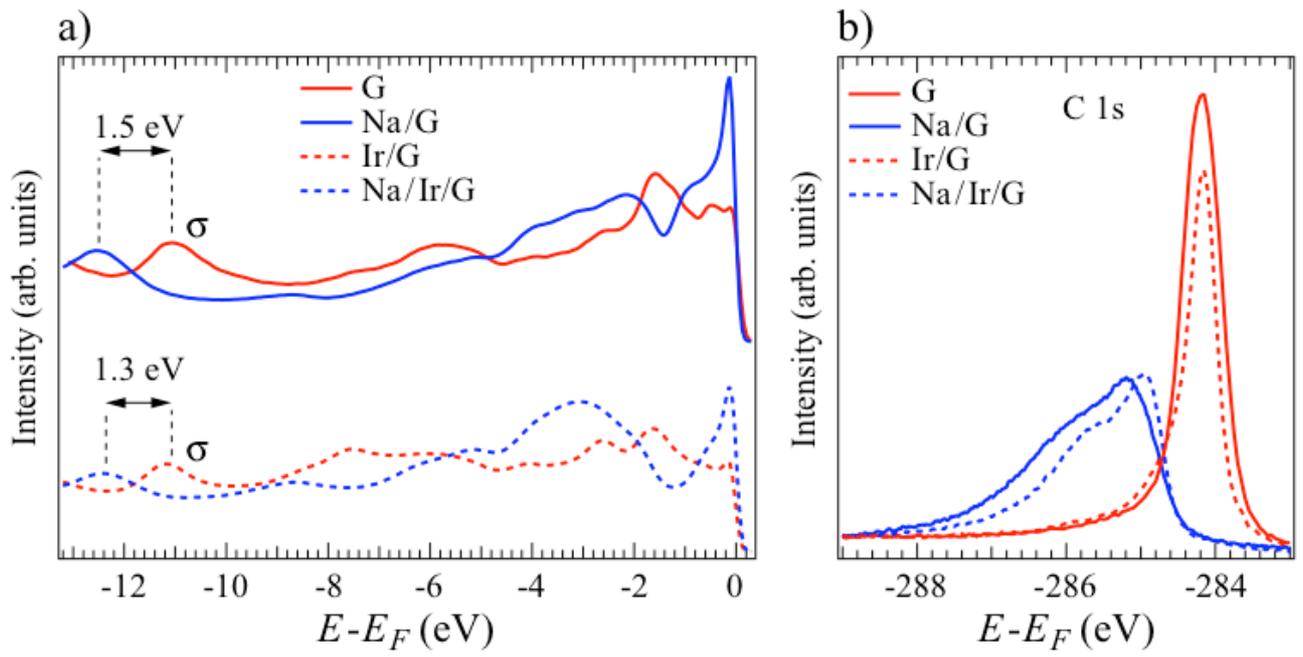

Figure 3

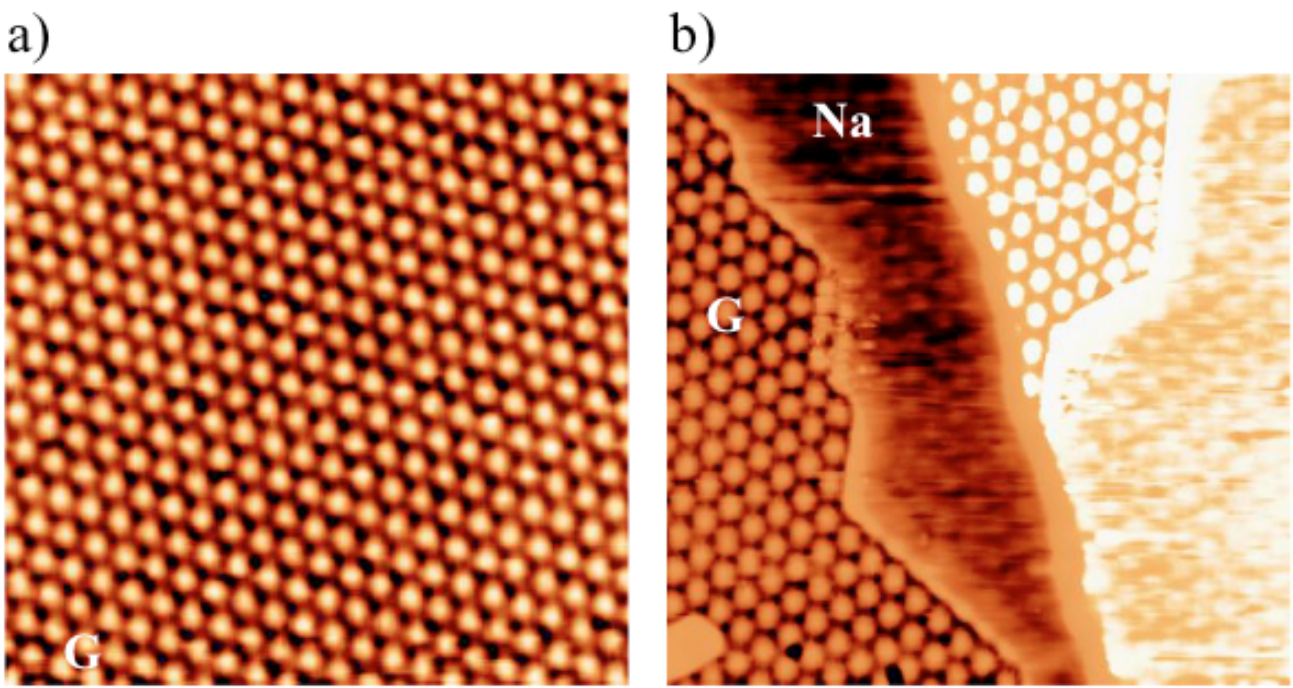

Figure 4

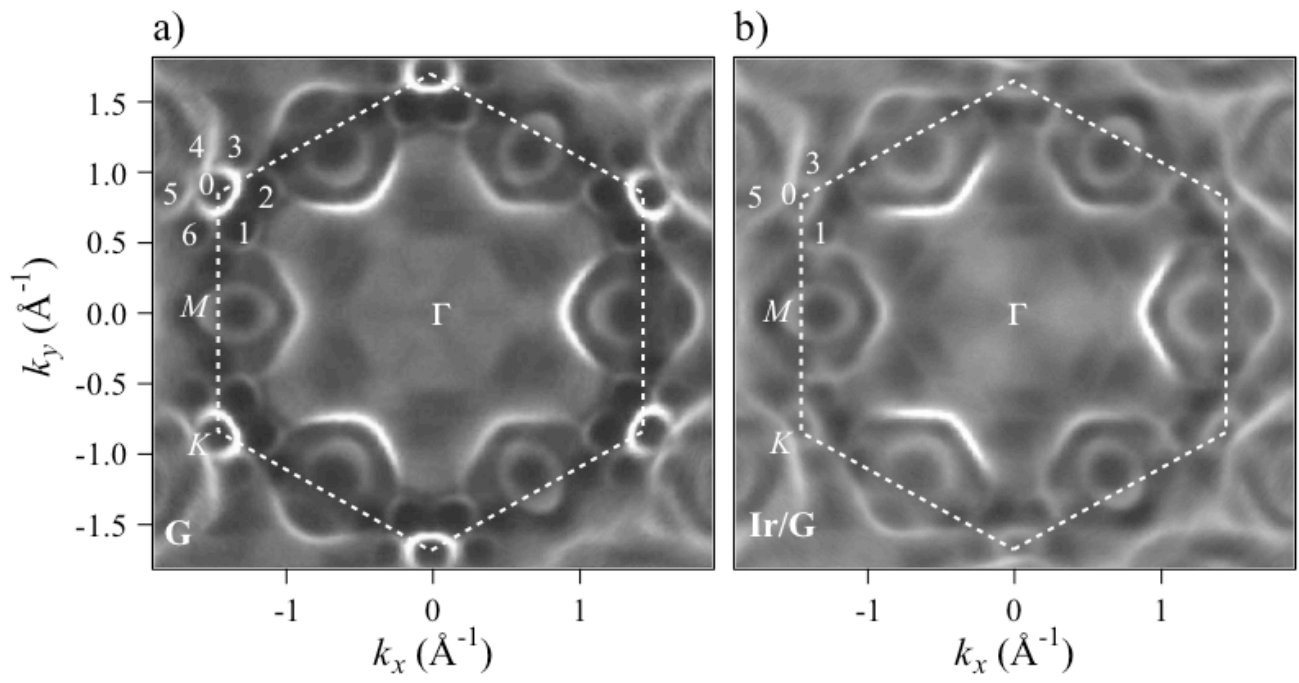

Figure 5